# Charge-tuneable biexciton complexes in monolayer WSe$_2$

Matteo Barbone[1,2,†], Alejandro R.-P. Montblanch[1,†], Dhiren M. Kara[1], Carmen Palacios-Berraquero[1], Alisson R. Cadore[2], Domenico De Fazio[2], Benjamin Pingault[1], Elaheh Mostaani[2], Han Li[3], Bin Chen[3], Kenji Watanabe[4], Takashi Taniguchi[4], Sefaattin Tongay[3], Gang Wang[2], Andrea C. Ferrari[2*], Mete Atatüre[1*]

[1]*Cavendish Laboratory, University of Cambridge, JJ Thomson Ave., Cambridge CB3 0HE, UK*
[2]*Cambridge Graphene Centre, University of Cambridge, Cambridge CB3 0FA, UK*
[3]*School for Engineering of Matter, Transport and Energy, Arizona State University, Tempe, AZ 85287, USA*
[4]*National Institute for Materials Science, Tsukuba, Ibaraki 305-0034, Japan*

Multi-exciton states such as biexcitons, albeit theoretically predicted, have remained challenging to identify in atomically thin transition metal dichalcogenides so far. Here, we use excitation-power, electric-field and magnetic-field dependence of photoluminescence to report direct experimental evidence of two biexciton complexes in monolayer tungsten diselenide: the neutral and the negatively charged biexciton. We demonstrate bias-controlled switching between these two states, we determine their internal structure and we resolve a fine-structure splitting of 2.5 meV for the neutral biexciton. Our results unveil multi-particle exciton complexes in transition metal dichalcogenides and offer direct routes to their deterministic control in many-body quantum phenomena.

In monolayer (1L) transition metal dichalcogenides (TMDs), the three-atom thickness of the material reduces the dielectric screening with respect to their bulk counterparts [1,2]. As a result of this and of the large effective mass, excitons (quasi-particle states formed of electrons and holes via Coulomb interaction) have binding energies of hundreds of meV [1,2] and are stable at room temperature. The physics of light-matter interaction is also enriched by two inequivalent valleys having opposite spin-locked valley indices [3] at the K points of the Brillouin zone, in which radiative recombination generates photons carrying opposite angular momenta [4,5]. These properties motivated the exploration of exciton and polariton [6] condensation [7,8] and superfluidity [9], and the exploitation of the spin and valley degrees of freedom as means to carry and manipulate information in quantum optoelectronic devices [3,10]. In the limit of quantum-confined excitons, the presence of localized single-photon emitters that can be induced deterministically [11,12] and generated by electroluminescence [13], makes TMDs a promising platform for the field of quantum photonics. Contrary to the exciton and trion states, optical studies of biexciton complexes in 1L-TMDs have been challenging [14–20]: inhomogeneous broadening [21] and defect bands [22] have limited their unambiguous identification and control. As a consequence, previous experimental findings [14–17,19,20] assigned neutral biexcitons with larger binding energy than that of the trions, in contrast to theoretical predictions [23–26], whereas Ref. 26 suggests neutral biexcitons in 1L-MoSe$_2$ in the consistent energy range.

In this work, we use recent advances in material and device processing [21,27] to suppress the effects that degrade the optical quality of 1L-WSe$_2$.

---
[†]These authors contributed equally to this work
[*] Corresponding authors A.C.F (acf26@cam.ac.uk) and M.A. (ma424@cam.ac.uk).

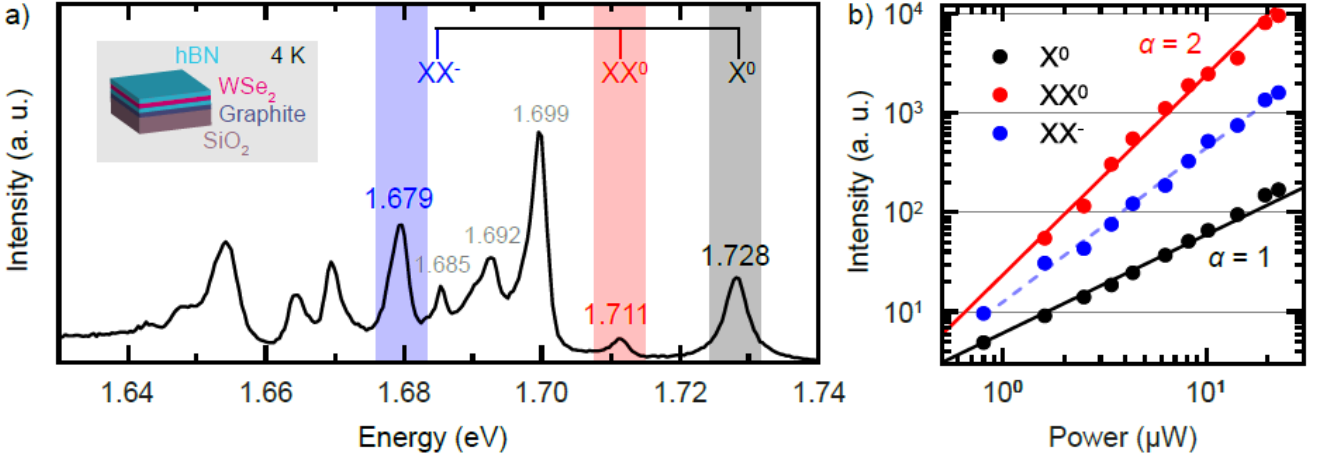

FIG. 1. PL spectrum and power dependence of encapsulated 1L-WSe$_2$ at 4 K. **(a)** PL spectrum (black curve, linear scale) of encapsulated 1L-WSe$_2$. Excitation wavelength: 658 nm. The top part of the figure lists the calculated spectral locations of X$^0$ (grey), XX$^0$ (red) and XX$^-$ (blue) transitions in the presence of a screening environment. **(b),** Double logarithmic plot of PL intensity as a function of excitation power for X$^0$ (black filled circles), XX$^0$ (red filled circles) and XX$^-$ (blue filled circles). The solid curves represent $I \propto P^\alpha$ for a quadratic ($\alpha = 2$, red) and linear ($\alpha = 1$, black) behaviour. The dashed blue curve is a fit to PL intensity, yielding an $\alpha$ of 1.55. For clarity of display, we multiply XX$^0$ by 4 and X$^0$ by 0.4.

To reduce the photoluminescence (PL) spectral linewidths [21] we place a layered material heterostructure (LMH) formed of 1L-WSe$_2$ encapsulated between two flakes of multilayer hexagonal boron nitride (ML-hBN) on a Si/SiO$_2$ substrate. To suppress the effect of SiO$_2$ charge traps we place a few-layer graphene (FLG) crystal below the bottom ML-hBN flake. The inset of Fig. 1a shows a schematic of the LMH (see Appendix for details on fabrication and characterisation).

We illuminate the LMH with continuous laser excitation at 658 nm and collect its optical emission at 4 K (see Appendix for further details on the optical measurements): Fig. 1a is a representative PL spectrum. Consistent with previous reports, we identify the bright neutral exciton [28], X$^0$, at ~1.728 eV, the negatively charged intervalley trion [29], X$^-_\text{inter}$, at ~1.699 eV, the negatively charged intravalley trion [29], X$^-_\text{intra}$, at 1.692 eV, and the dark neutral exciton [30,31], X$^0_\text{dark}$, at ~1.685 eV. Here, *bright* refers to excitons with in-plane dipole and spin-allowed radiative recombination [2,30,31], whereas *dark* refers to excitons with out-of-plane dipole and spin-forbidden radiative recombination [2,30,31], for which emission only occurs in plane but is captured partially by our high numerical aperture objective. The peak at ~1.711 eV is a good candidate for a neutral biexciton (XX$^0$), as it appears in the predicted energy range [23–26]. The peak at ~1.679 eV was previously labelled as neutral biexciton emission [14], although it appears in the energy range predicted for the negatively charged biexciton (XX$^-$) [24,26]. In the top part of Fig. 1a, we include the emission energies of single- and multi-exciton species in ML-WSe$_2$ calculated via the diffusion Quantum Monte-Carlo [26] technique combined with environment screening (see Appendix).

Figure 1b displays the PL intensity $I$, defined as peak area, as a function of excitation power $P$ (with $I \propto P^\alpha$) for X$^0$ (filled black circles), XX$^0$ (filled red circles) and XX$^-$ (filled blue circles). For reference, we plot solid curves corresponding to a linear ($\alpha = 1$, black) and quadratic ($\alpha = 2$, red) behaviour. We expect superlinear behaviour for biexcitons reaching $\alpha = 2$ in the thermodynamic equilibrium [32][,11]. The power dependence of XX$^0$ follows the quadratic curve, while that of XX$^-$ is superlinear with fitted $\alpha \sim 1.55 \pm 0.03$ (dashed blue curve). Both trends of XX$^0$ and XX$^-$ are therefore consistent with a biexcitonic origin and contrast the linear behaviour of $X^0$. The deviation of $XX^-$ from $\alpha = 2$ possibly stems from the competition of electron capture from other optically induced excitons. Remaining peaks of Fig. 1a follow an approximately linear power dependence.

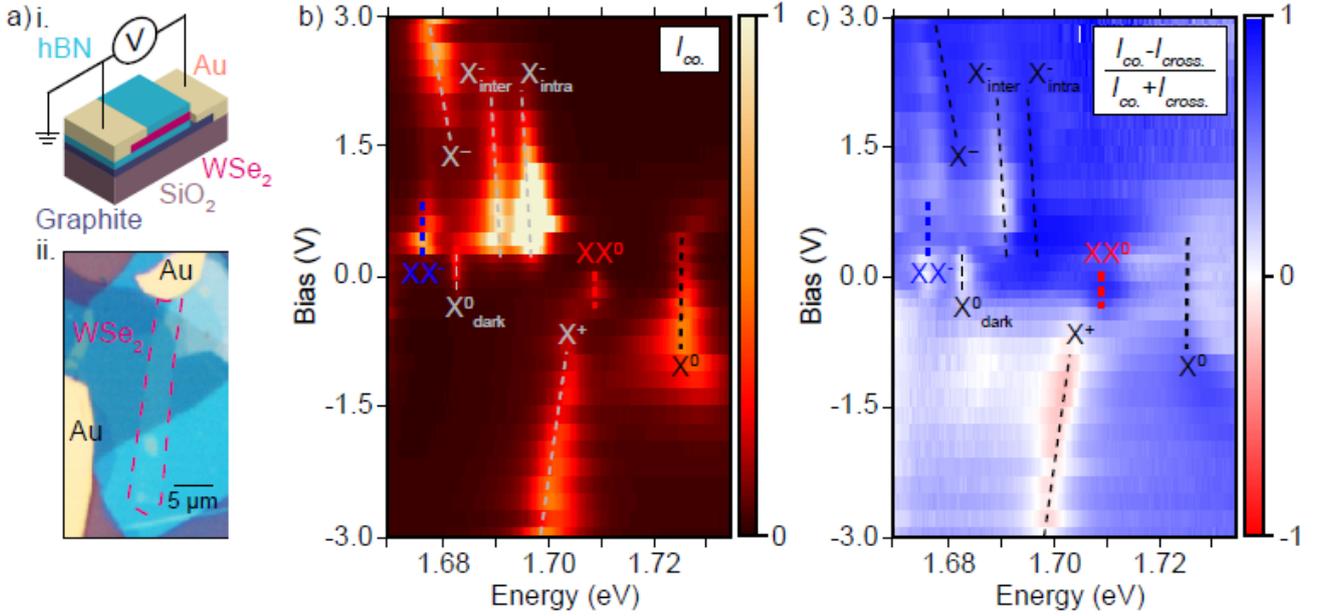

FIG. 2: Charge dependence of PL. **(a)** (i) Schematic and (ii) optical image of the charge-tuneable device. The red dashed frame highlights the 1L-WSe$_2$ flake. **(b)** Circular co-polarised PL intensity ($I_{\sigma+/\sigma+}+I_{\sigma-/\sigma-}$) as a function of applied bias. The dashed lines are a guide to the eye to highlight each peak. **(c)** DoP of PL as a function of bias and energy in the same range as panel **b**. The colour code is such that blue regions indicate co-polarisation, whereas the red regions indicate counter-polarisation with respect to excitation polarisation.

To differentiate the charged and neutral biexciton $XX^0$ and $XX^-$, we fabricate a charge-tuneable device starting from a new LMH analogous to the first one but with the addition of one electrode to the FLG and of a second electrode to an uncovered 1L-WSe$_2$ portion (see Methods). The subpanels of Fig. 2a show the schematic and the optical image of the device. Figure 2b displays how the PL spectrum is modified as a function of voltage $V$. The charging regime modifies the optical signatures of 1L-WSe$_2$ strongly. The presence of $X^0$ and $X^0_{dark}$ at $V \sim 0$ V shows that the material has a negligible intrinsic charge doping. At the same bias, Fig. 2b also shows emission from $XX^0$. In the electron-charged regime ($V > 0$) fluorescence from $X^0$, $XX^0$ and $X^0_{dark}$ vanishes, while emission from $X^-_{inter}$, $X^-_{intra}$ and $XX^-$ arises. Around 2 V the $X^-$ emission switches to a new peak at ~1.681 eV, likely the next charging state of the trion, $X^{--}$. This peak was previously assigned to the fine structure of $X^-$ on experiments on bare material [10]. Negative bias is the hole-charged regime, where only $X^0$ and the positively charged trion $X^+$ are visible (Refs [10,29]).

We then analyse the correlation between excitation and emission polarisations in the different charging regimes (Fig. 2c). We plot the degree of circular polarization [DoP = ($I_{co.}$-$I_{cross.}$) / ($I_{co.}$+$I_{cross.}$) [4]] where $I_{co.}$ ($I_{cross.}$) is the intensity of the circularly polarized light with the same (opposite) helicity in the excitation and detection paths. We refer to the two orthogonal helicities as σ- and σ+. At 0 V, $XX^0$ has DoP > 80%, while $X^0_{dark}$ shows no circular polarization [30,31], as expected. At 0.8 V, $X^-_{inter}$ has DoP > 90%, $X^-_{intra}$ has DoP < 10% and $XX^-$ has DoP ~ 55%. The circular polarization of photons from both $XX^0$ and $XX^-$ thus implies that dissociation occurs with the recombination of a bright exciton, as a dark exciton would emit linearly polarized light [30,31]. The DoP of $XX^-$ is close to the average of the DoP of $X^-_{inter}$ and $X^-_{intra}$, suggesting that the recombination mechanisms of both $X^-_{inter}$ and $X^-_{intra}$ contribute [33] to that of $XX^-$.

The electrons and holes comprising the biexcitons can occupy multiple combinations of band states. To identify them, we resort to the variation of PL as a function of an out-of-plane magnetic field. Figure 3a, panel i, shows the σ- polarized PL of $X^0$ and $XX^0$ under co-polarized (σ-) excitation. We resolve a finite splitting in the $XX^0$ emission, with a separation of 2.5 meV between the two peaks labelled $XX^0_1$ and $XX^0_2$ (line-cut spectra at different magnetic fields are shown in Appendix, Fig. A3). This fine-structure splitting is likely introduced by exchange interaction, in analogy to the case of the splitting between $X^-_{inter}$ and $X^-_{intra}$ [29,33,34]. This experimental observation of the $XX^0$ fine structure will set a reference for further computational studies, which otherwise suffer from limitations due to the complex treatment of the exchange interaction. Figure 3a, panel ii, shows the σ+ polarized PL of $X^0$ and $XX^0$ under cross-polarized (σ-) excitation. Here, only $XX^0_2$ remains visible, revealing a different DoP for $XX^0_1$ and $XX^0_2$, in analogy to the different DoP between $X^-_{inter}$ and $X^-_{intra}$. Additionally, the PL intensity of $XX^0$ emission increases when it shifts to higher energies, in contrast to that of $X^0$. We observe the same behaviour for $XX^-$ in Fig. 3b, where the co-polarized PL from the recombination of the quasi-particle also shows valley-dependent Zeeman shift.

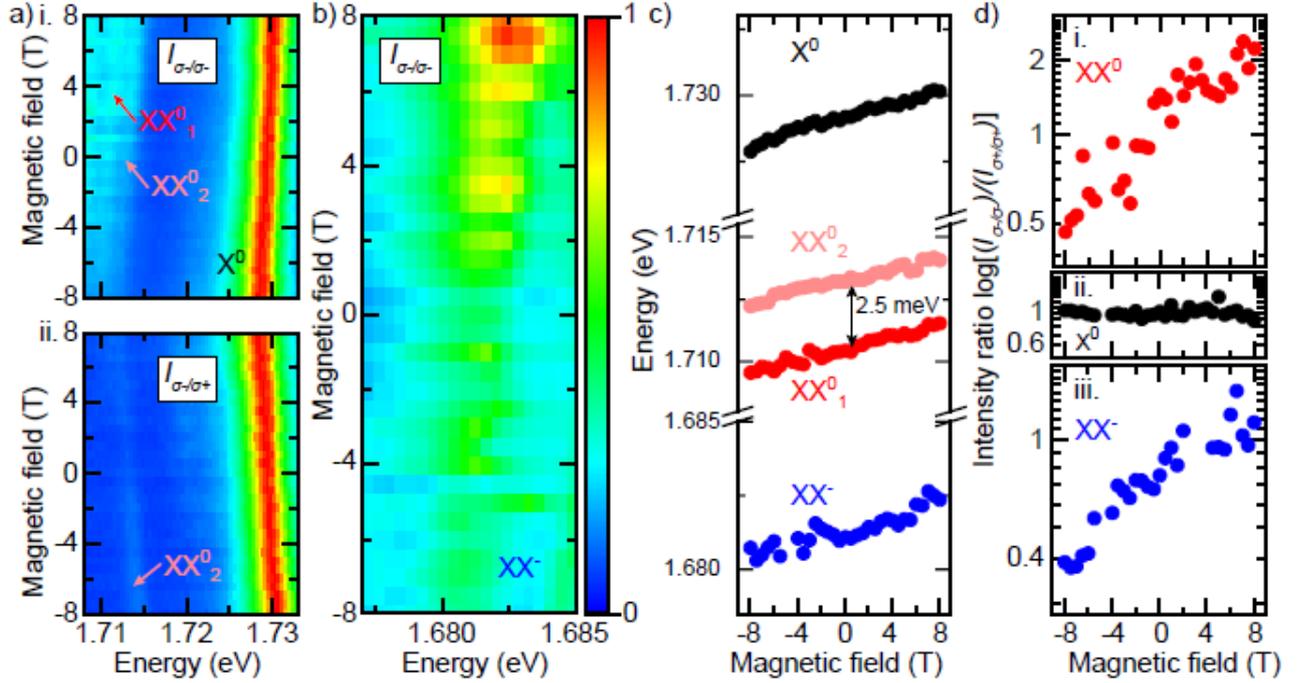

FIG. 3: Magnetic field dependence of PL. **(a)** Magnetic field dependent PL of $X^0$ and $XX^0$ in (i) circular co-polarised and (ii) cross-polarised configurations, for σ- excitation. The fine-structure lines are indicated as $XX^0_1$ and $XX^0_2$. The emission of $XX^0$ brightens with increasing emission energy. $X^0$ is displayed for reference. **(b)** Magnetic field dependent PL of $XX^-$ in a circular co-polarized configuration, for σ- excitation. In **a** and **b** the colour scale is linear. **(c)** Zeeman shift in the PL spectrum of $X^0$ (filled black circles), $XX^0$ (filled red and pink circles for the two components of the fine-structure) and $XX^-$ (filled blue circles). The double arrow is a scale bar of 2.5 meV. **(d)** PL intensity ratio of circular co-polarization with opposite helicity $I(σ-/σ-)/I(σ+/σ+)$ for $X^0$, $XX^0_1+XX^0_2$ and $XX^-$ as a function of magnetic field.

In Fig. 3c we plot the energies of $X^0$, $XX^0_1$, $XX^0_2$ and $XX^-$ as a function of magnetic field. For each multi-exciton species, we calculate the Landé factor $g$, defined as $ΔE = gμ_B B$, where $ΔE = E_{σ+} - E_{σ-}$ is the difference in the emission energy of excitons in opposite valleys, $μ_B = eℏ/2m_e$ = 58 μeV/T is the Bohr magneton and $B$ is the magnetic field. We derive $g$ ~ -4.44 ± 0.12 for $X^0$, consistent with previous observations [35], ~ -4.10 ± 0.15 for $XX^0$ and ~ -3.86 ± 0.17 for $XX^-$.

The emission intensities of $XX^0$ and $XX^-$ change dramatically with magnetic field, being stronger when shifted to higher energy. Figure 3d displays the $I_{σ-/σ-}/I_{σ+/σ+}$ ratio as a function of magnetic field for $XX^0_1+XX^0_2$ (panel i) and $XX^-$ (panel iii). For comparison, we show $I_{σ-/σ-}/I_{σ+/σ+}$ for $X^0$ in panel ii. At zero magnetic field $I_{σ-/σ-}/I_{σ+/σ+}$ is ~1 for all peaks, i.e. the two valleys have the same exciton population.

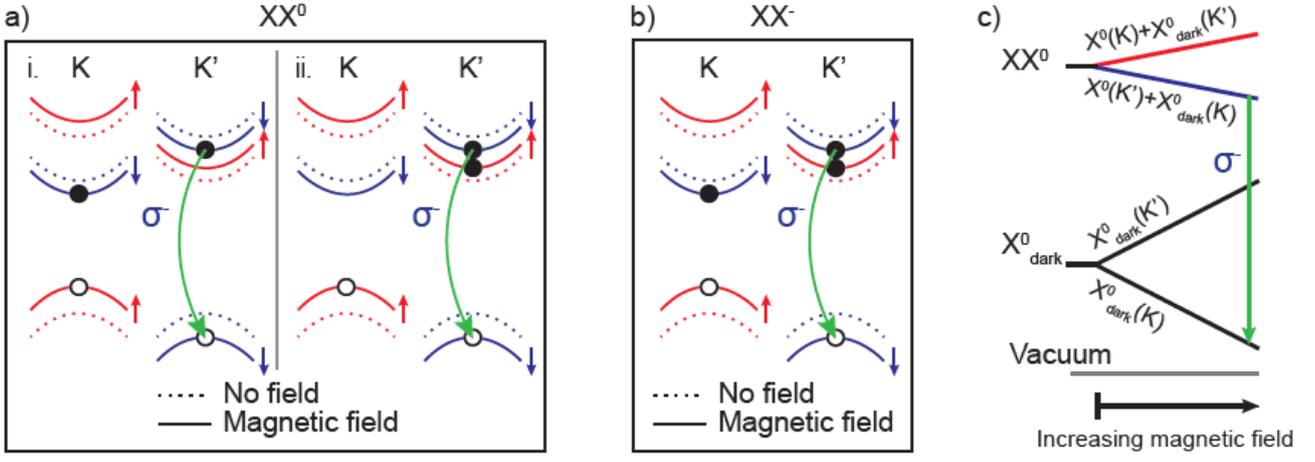

**Figure 4: Composition of biexciton species with applied magnetic field. (a,b)** Single-particle picture of the internal structure of (**a**) $XX^0$ and (**b**) $XX^-$ for B > 0. The eigenstates shift inequivalently in K and K' (dashed curves indicate no magnetic field, solid curves indicate applied magnetic field, red and blue colours indicate opposite spin). $XX^0$ comprises a bright exciton with highest radiative energy and a dark exciton with the electron (**a** panel i) inter- or (**a** panel ii) intra-valley with the bright exciton. (**c**) Many-body picture of the magnetic field effect on $XX^0$, comprising a bright and a dark exciton. Applying a magnetic field shifts the energy of the dark exciton more than that of $XX^0$ due to the higher g factor of the former. This results in the dissociation of the biexciton in the form $XX^0 \rightarrow X^0_{dark}+\gamma(\sigma\text{-})$, where $\gamma(\sigma\text{-})$ is a photon with σ- helicity.

When magnetic field is applied, $I_{\sigma\text{-}/\sigma\text{-}}/I_{\sigma+/\sigma+}$ remains unaffected for $X^0$ (Fig. 3d, panel ii). This can be explained by $X^0$ in each valley recombining before reaching thermal equilibrium. In stark contrast, $XX^0$ and $XX^-$ display strongly anti-symmetric magnetic-field dependence, as shown in Fig. 3d, panels i and iii: for increasing magnetic field, the lower-energy transition is weaker.

We can understand the complex behaviour of the magnetic-field dependent PL through the single-particle picture of the energy bands. Figures 4a,b illustrate the effect of $B > 0$ on the band structure of 1L-WSe$_2$ around the K and K' points, considering the contribution of the spin and atomic orbital magnetic moments. The 1L-WSe$_2$ bandgap decreases (increases) in the K (K') valley as the energies of both hole and electron experience the same spin upshift (downshift), while the hole experiences a larger orbital upshift (downshift) [35,36] with respect to the electron. The applied magnetic field induces anti-symmetric exciton populations in the two valleys (Fig. 3d). This excludes the possibility that $XX^0$ (Fig. 4a) may be formed by two bright or two dark excitons, as both cases would result in equally intense radiative recombination from both K and K' at all magnetic fields. $XX^0$ is therefore a combination of a bright and a dark exciton, with the bright exciton of each valley locked to the opposite photon helicity [3]. Under magnetic field, the bright exciton component of $XX^0$ occupies the higher-radiative energy transition (Figs 3a,d) due to thermalization of the photogenerated electrons to the K' valley. This is allowed by the considerably long lifetime of XX, namely ~2 to 100 times longer than single excitons [14,19]. In parallel, the electron of the dark component of $XX^0$ can be either in the opposite (Fig. 4a, panel i) or in the same (Fig. 4a, panel ii) valley as the bright exciton component, yielding an energy shift between these two configurations, which is the origin of the fine-structure of $XX^0$ observed in Fig. 3c.

Figure 4b illustrates the single-particle configuration of $XX^-$. As for $XX^0$, the combination of two bright excitons is excluded due to different recombination intensities in K and K'. From the similar $g$ of $XX^0$ and $XX^-$, we can understand this five-particle complex as a bound state of a bright exciton with a dark trion, or a bright trion with a dark exciton. Both configurations would show inequivalent valley population as for $XX^0$ in Fig. 3d.

Figure 4c is a qualitative many-body picture for $XX^0$ formed by a bright and a dark exciton component in opposite valleys under magnetic field. As its total Zeeman splitting depends on both the bright and the dark component, $XX^0$ splits with a reversed energy order compared to its bright exciton component and dissociates into a dark exciton and a photon due to the dark exciton having larger $g$ than $X^0$ with opposite sign [37]. The distribution of biexciton states follows the case near thermal equilibrium, which is the reason behind the inequivalent circularly co-polarised emission intensity under σ+ or σ-, as shown in Fig. 3.

We have discovered the five-particle negatively charged biexciton in 1L-WSe$_2$, as well as the neutral biexciton resolving its fine structure Immediate next steps include the unequivocal verification of the $X^{--}$ state and the identification of bound states within the lower-energy peaks. A complete understanding of multi-exciton complexes is key to study coherent many-body phenomena, such as condensation [7,8] and superfluidity [9]. Further, the ability to access and manipulate biexciton complexes in TMD-based heterostructures offers new routes towards probing other fundamental excitations in this system and the interplay between free and localized excitons. Extending our findings to the quantum confined regime will open new capabilities for cascaded emission of entangled photons and spin-multiphoton interfaces.

We thank Bernhard Urbaszek, Neil D. Drummond and Vladimir I. Fal'ko for useful discussions. We acknowledge funding from NSF DMR-1552220, Elemental Strategy Initiative conducted by the MEXT, Japan and the CREST (JPMJCR15F3), JST, EU Graphene Flagship, ERC Grants Hetero2D and PHOENICS, EPSRC Grants EP/509K01711X/1, EP/K017144/1, EP/N010345/1, EP/M507799/ 5101, and EP/L016087/1, Marie Skłodowska-Curie Actions Spin-NANO, Grant No. 676108, Quantum Technology Hub NQIT EP/M013243/1.

**APPENDIX**

**1. Materials sourcing, characterization and device assembly**

W and Se pellets with ultra-high purity (Puratronic 99.9999% or higher) are sealed in a quartz ampoule at $10^{-6}$ Torr pressure. We adopt a two-step growth approach: in the first step, W and Se pellets are sealed at stoichiometric ratios (~500 mg total weight) in a quartz ampoule. The pressure is kept at $10^{-6}$ Torr and the temperature at 1050 $^{0}$C, slightly below the actual growth temperature of 1065 $^{0}$C. Before sealing in vacuum, T~300 $^{o}$C is applied to remove any residual molecules adsorbed on the precursors as well as on the quartz reactor walls. After sealing, the ampoule is at 1050 $^{o}$C for 1 week to produce polycrystalline WSe$_2$ powders. The WSe$_2$ powders are checked with scanning electron microscope-energy dispersive spectroscopy (SEM-EDS) to determine the amount of Se vacancies. In the second step, ~5 mg extra Se is sealed with the WSe$_2$ powders in a quartz ampoule at $10^{-6}$ Torr. The excess Se helps reducing potential Se vacancies. The sealed ampoule is then heated to 1090 $^{o}$C over 3 days. The side without precursors is cooled by 25 $^{o}$C to 1065 $^{o}$C within an hour to create the thermodynamic flux to initiate the growth [38]. This process typically leads to crystalline WSe$_2$ flakes that are a few mm in size. Bulk hBN crystals are grown by the temperature-gradient method under high pressure and high temperature described in Ref. 40. Graphite is sourced from NGS Naturgrafit. All bulk crystals are exfoliated by micromechanical cleavage [39] on Si/SiO$_2$ (oxide thickness 285 nm). 1L- and FL samples are identified by optical contrast [40]. Selected crystals are assembled within ~5 hours into LMHs via dry-transfer [27] as detailed in Ref. [13].

The LMH sample used for power-dependent and magnetic field dependent PL measurements is formed, from top to bottom, of ML-hBN flakes (~5 nm thick as determined by optical contrast), 1L-WSe$_2$, and a second ML-hBN flake (~10 nm thick as determined by optical contrast) optical contrast) and FLG (~ 5

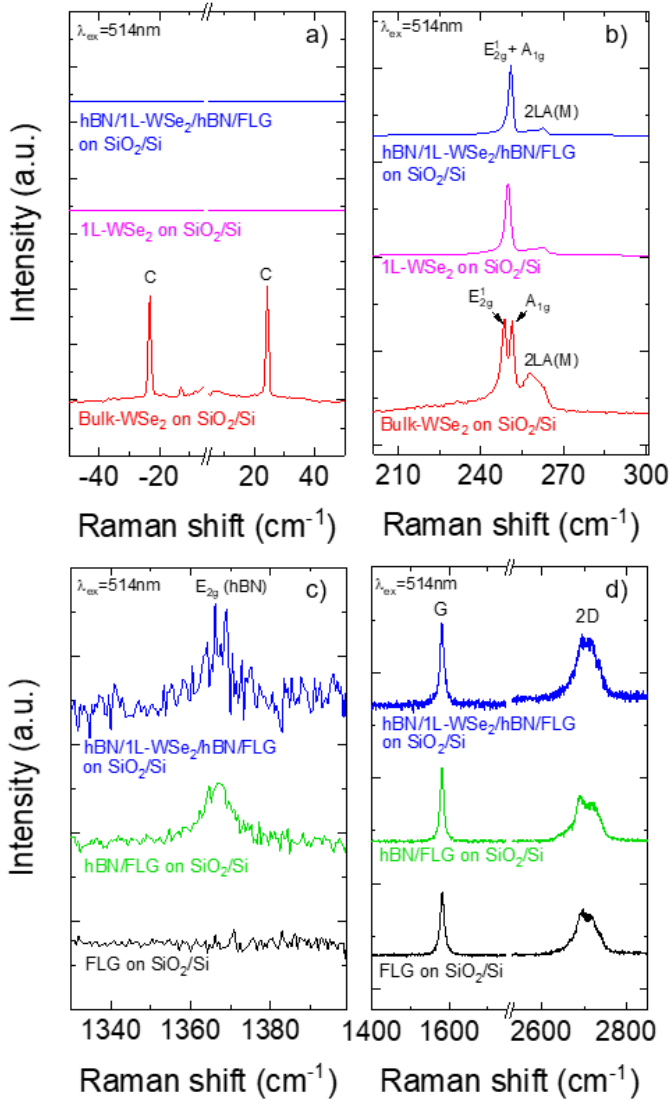

FIG. A1: Raman spectra. a) Low frequency and b) high frequency Raman spectra of Bulk $WSe_2$ (red), 1L-$WSe_2$ (magenta) on Si/$SiO_2$ and ML-hBN/1L-$WSe_2$/ML-hBN/FLG (blue). c) $E_{2g}$ peak of hBN and D peak spectral regions and d) G and 2D peaks of FLG spectral region in FLG on $SiO_2$ (black); ML-hBN/FLG (green) on Si/$SiO_2$ and ML-hBN/1L-$WSe_2$/ML-hBN/FLG (blue).

layers thick as determined by optical contrast). That used for voltage-dependent measurements is prepared in a similar way, but the top ML-hBN does not fully cover the 1L-$WSe_2$ to allow for Cr/Au (5/50 nm) electrodes to directly contact it. The second electrode contacts FLG. The electrodes are patterned by e-beam lithography followed by lift-off. The ML-hBN thickness is chosen to isolate the 1L-$WSe_2$ from the environment, smoothen the roughness of $SiO_2$, shield the charge-traps of the substrate and avoid tunnelling between FLG and 1L-

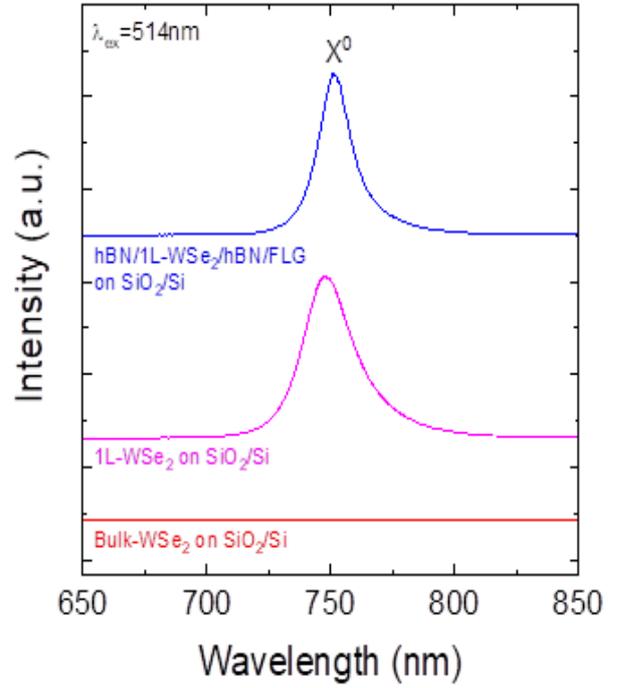

FIG. A2 PL spectra at RT. Blue curve shows the $X^0$ exciton at ~750 nm, which confirms the presence of 1L-$WSe_2$ [41]. The red curve is the PL spectrum of bulk $WSe_2$ on $SiO_2$/Si measured keeping the same laser power as that of 1L-$WSe_2$ (magenta curve). The blue curve is the PL spectrum of the LMH.

$WSe_2$, while not compromising the optical contrast under the optical microscope.

## 2. Room-temperature optical characterization

The LMHs are characterized by Raman spectroscopy and PL using a Horiba LabRAM Evolution spectrometer equipped with a 100x objective (N.A. 0.6). Fig. S2.1 (a,b) plot the spectrum of bulk (red curve) and 1L-$WSe_2$ (magenta curve). In 1L-$WSe_2$, the $E_{2g}^1$ and $A_{1g}$ modes [41] are merged in a single band at ~250 cm$^{-1}$ and the shear mode C, due to relative motion of atoms in adjacent planes, is not present [41]. Bulk $WSe_2$ shows split $E_{2g}^1$ and $A_{1g}$ modes at ~249 cm$^{-1}$ and at ~251 cm$^{-1}$ [41]. The C mode is detected at ~24 cm$^{-1}$ [42]. Fig. S2.1 (c,d) plot the Raman spectra of the resulting LMH, corresponding to Fig. 2 in the main text. The black curve in Fig S2.1 (c,d) is the spectrum of FLG on $SiO_2$/Si with the characteristic G peak at 1580 cm$^{-1}$.

PL (Fig. S2.2) is collected to confirm the monolayer nature of the $WSe_2$ crystal. The peak

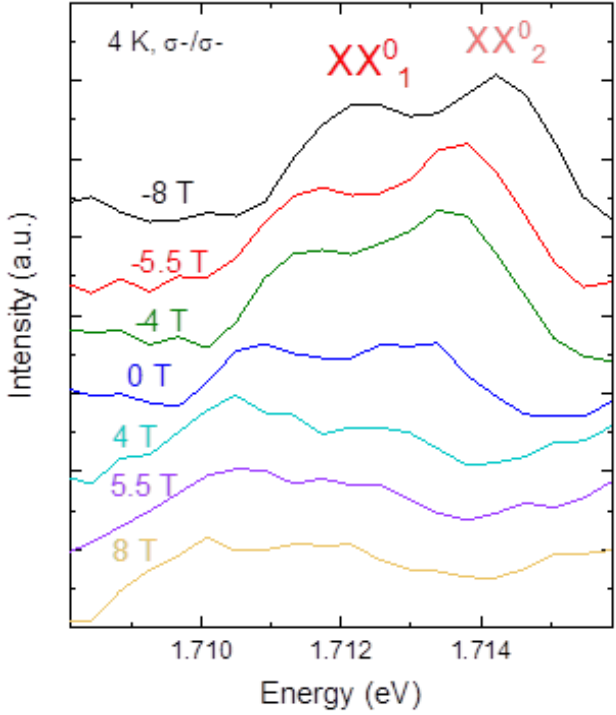

FIG. A3: Line-cut spectra of $XX^0$ as function of applied magnetic field. Spectra are taken at 4 K and different magnetic fields with a circular co-polarized scheme and σ- helicity, showing the double peak formed by $XX^0_1$ and $XX^0_2$, after linear background subtraction.

at ~750 nm corresponds to the neutral $X^0$ exciton [41].

Fig S2.2 compares the PL spectrum of 1L-$WSe_2$ on $SiO_2$ (magenta) and in the LMH (blue) with that of bulk $WSe_2$ (red) keeping the same measurement conditions. The PL intensity drastically decreases in bulk $WSe_2$ with respect to 1L-$WSe_2$ due to a direct-to-indirect bandgap transition[7]. The shape of the 1L-$WSe_2$ PL is preserved in the LMH.

**3. Optical measurements at 4 K:** Power dependent and gate-controlled measurements are performed in a variable-temperature Helium flow cryostat (Oxford Instruments Microstat HiRes2) with a home-built confocal microscope at a nominal temperature of 4.2 K. The magneto-optical measurements are performed in a close-cycle bath cryostat (Attocube Attodry 1000) equipped with a superconducting magnet (maximum out-of-plane magnetic field 8 T) at a nominal sample temperature of 3.8 K. Thus, in the main text we refer to measurements at 4K, as an average of these 2 nominal temperatures. Fig. A3 shows line-cut spectra of the fine-structure of $XX^0$ for different magnetic fields.

**4. Theoretical Calculations**

We use Mott-Wannier model and quantum Monte Carlo (QMC) as implemented in CASINO [43] to calculate the energy of $X^0$, $XX^0$ and $XX^-$ in ML-$WSe_2$ [26]. The full photoemission spectra of ML-$WSe_2$ in vacuum is reported in Ref. [26]. To consider the effect of the dielectric screening provided by hBN, we use the experimental value of the binding energy of $XX^0$ and use Eq. 48 of Ref. [26] to derive the screening parameter $r^*$ which is 54 Å. We also use the many-body GW electron and hole effective masses as 0.29 $m_0$ and 0.34 $m_0$ [44] respectively, where $m_0$ is the bare electron mass. Then we calculate the binding energy of $XX^-$ by subtracting the total energy of the exciton and trion from the total energy of $XX^-$.